\documentclass[slac_one]{revtex4}
\usepackage{graphicx}
\usepackage{fancyhdr}
\usepackage{comment}
\begin{document}

\title{Combination of Single Top Quark Production Results from CDF} 

\author{R. C. Group (on behalf of the CDF Collaboration)}
\affiliation{FNAL, Batavia, IL 60510, USA}

\author{C.I.~Ciobanu}
\affiliation{LPNHE, Universite Pierre et Marie Curie,Paris, F-75252, France}

\author{K.~Lannon}
\affiliation{The Ohio State University, Columbus, OH  43210, USA}

\author{C.~Plager}
\affiliation{UCLA, Los Angeles, CA  90024 USA}

\begin{abstract}
Recently, the CDF experiment at the Fermilab Tevatron has used
complementary methods to make multiple measurements of the singly produced
top quark cross section.  All analyses use the same dataset with more than
2 fb$^{-1}$ of CDF data and event selection based on $W+2$ or $W+3$ jet
events with at least one $b$-tagged jet.  However, due to differences in
analysis techniques these results are not fully correlated and a
combination provides improved experimental precision.  Two independent
methods are used to combine the results. This combination
results in an improved measurement of the single top production cross section and
also the CKM matrix element $V_{tb}$.
\end{abstract}

\maketitle

\thispagestyle{fancy}

\section{Single Top Results from CDF} 

Measuring the properties of single top production at the Tevatron is
challenging.  The production cross section is smaller than top pair
production and the final state is harder to distinguish from major
backgrounds.  The background rates are large; in fact, the
error on the background prediction is larger than the predicted single
top signal.  In order to improve separation of signal from background
three multivariate techniques were used at CDF: a likelihood function
(LF), a matrix element (ME), and a neural network (NN)~\cite{STPRL}.

In each analysis the multivariate discriminant was used to build
template distributions for signal and background expectations and
the data was fit to extract the signal component.  Although the
analyses use the same event selection, they rely on different
observables to discriminate signal from background.  Although many of
these observables are highly correlated, a combination of results
should provide additional sensitivity.  Two very different strategies
were developed to combine the CDF single top analyses.

\section{NEAT Combination}
\begin{figure*}[t]
  \centering
  \includegraphics[width=65mm]{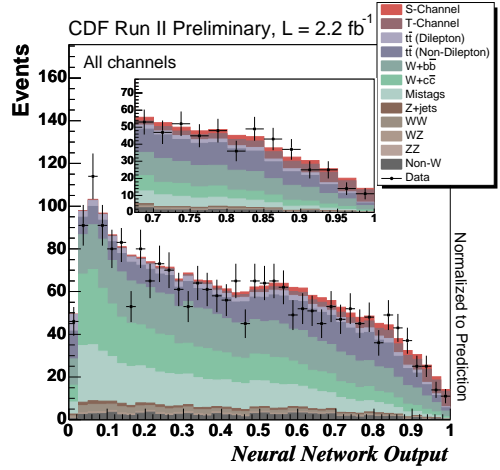}
  \caption{NEAT output of the background prediction and CDF data.}
  \label{fig:NEAT_stack}
\end{figure*}

The NEAT combination method takes the discriminating variable from
each analysis and combines them into one ``super-analysis''.  The new
super-discriminant is a neural network in which the
weights and topology are optimized for sensitivity by using a technique
known as neuro-evolution of augmenting topologies (NEAT)~\cite{neat}.  A separate discriminant is 
optimized for events with two or three jets and events with one or two
$b$-tags.  As in the individual analyses, templates based on the
event-by-event output of NEAT are built for signal and background
expectations and the data is fit to these to extract the most likely
single top component.  Figure~\ref{fig:NEAT_stack} shows the expected composition of signal
and background as well as the CDF data for all tag and jet channels combined.

\section{BLUE Combination}
As a cross check to the NEAT combination, the
method of forming a Best Linear Unbiased Estimate, BLUE, is used to
make a weighted average of the three single top cross section
measurements~\cite{blue1}.  Recall a simple $\chi^2$ combination
method:
\begin{equation}
  \label{eqn:chi2_1}
  \chi^2(\mu) = \sum_i^N \left( \frac{ m_i - \mu } {\sigma_i} \right)^2
  \quad \textrm{or} \quad 
  \chi^2(\mu) = {\bf \delta}^T(\mu) \cdot {\bf S}^{-1} \cdot {\bf
    \delta} (\mu)  
\end{equation}

  where $m_i$ and $\sigma_i$ are the measured value and error of the
 {\em i}th analysis, $\mu$ is the average of the measurements,
 $\delta(\mu)$ is the column vector $\delta_i = m_i - \mu$, and ${\bf
 S}$ is the covariance matrix.  In this case, ${\bf S}$ is diagonal
 with ${\bf S}_{(i,i)} = \sigma_i^2$.  BLUE is based upon the formula
 above, but in order to combine $N$ measurements which each have $E$
 different types of errors, ${\bf S}$ is defined as 
 \begin{equation}
 {\bf S} = \sum_{e}^E {\bf S}_e
\quad \textrm{and} \quad 
{{\bf S}_e}_{(i,j)} = {\sigma_e}_i \cdot {\sigma_e}_j 
 \cdot {\rho_e}_{(i,j)}
 \end{equation}
 where ${\sigma_e}_i$ is the $e$th type of error on the $i$th
measurement and ${\rho_e}_{(i,j)}$ is the correlation of the $e$th
error between the $i$th and $j$th measurement.

The beauty of BLUE lies in the fact that a minimization routine on
this $\chi^2$ need not be run to get the mean.  After defining ${\bf
  H}$ to be the inverse of the covariance matrix and $sum_H$ as below,
a weight for each measurement, $w_i$, can then be calculated: 

 \begin{equation}
   \quad \textrm{Define} \quad
 {\bf H} \equiv {\bf S}^{-1}
\quad \textrm{and} \quad
 sum_H \equiv \sum_{i=1}^N \sum_{j=1}^N {\bf H}_{i,j}
\quad \textrm{so that} \quad
w_i \equiv \sum_{j=1}^N \frac {{\bf H}_{i,j}}{sum_H}\\
\end{equation}
BLUE then predicts 
\begin{equation}
 \mu_\textrm{best} = \sum_{i=1}^N w_i \cdot m_i
\quad \textrm{and} \quad
\sigma = \frac{1}{\sqrt{sum_H}}
 \end{equation}
 where $\mu_\textrm{best}$ is the minimum of the $\chi^2$ and $\sigma$
 is the total error.

BLUE needs the measured values, uncertainties, and correlations
between each analysis before the combined result can be calculated.
Correlated pseudo-experiments are thrown from fully simulated events
matched between each analysis and the correlations between analyses of
the resulting cross sections measured in each pseudo-experiment are
summarized in Table~\ref{tab:corr}. 

\begin{table}[htb]
\begin{center}
\begin{tabular}{|l|c|c|c|}\hline
              & LF   & ME    & NN \\ \hline
LF (1.79~pb)  & 1.0  & 0.599 & 0.741\\ \hline
ME (2.17~pb)  & ---  & 1.0   & 0.609 \\ \hline
NN (1.97~pb)  & ---  & ---   & 1.0 \\ \hline 
\end{tabular}
\caption[]{Correlations coefficients between pairs of cross section values.}
\label{tab:corr}
\end{center}
\end{table}

In general, errors may have dependence on the value measured.
Analyses which measure lower than expected could have a smaller error
and this may bias the combination.   In order to avoid this bias we
run BLUE iteratively.   Another complication is that errors need
not be symmetric.  Since BLUE is based on a Gaussian approximation it
is not able to treat this directly.   To handle this we use Asymmetric
Iterative BLUE (AIB).  AIB is a set of three BLUE combinations which
uses average errors (for central value), upper errors, or lower errors
as follows:

\begin{equation}
{\mathcal R}_\textrm{upper} = \frac{\sigma_\textrm{upper BLUE}}{\sigma_\textrm{upper BLUE} + \sigma_\textrm{lower BLUE}}
\end{equation}

\begin{equation}
\sigma_\textrm{upper} = 2 \cdot {\mathcal R}_\textrm{upper} \cdot \sigma_\textrm{center BLUE} 
\quad \textrm{and} \quad
\sigma_\textrm{lower} = 2 \cdot (1 - {\mathcal R}_\textrm{upper}) \cdot \sigma_\textrm{center BLUE} 
\end{equation}

\subsection{Results}
\begin{figure*}[ht]
  \centering
  \includegraphics[width=135mm]{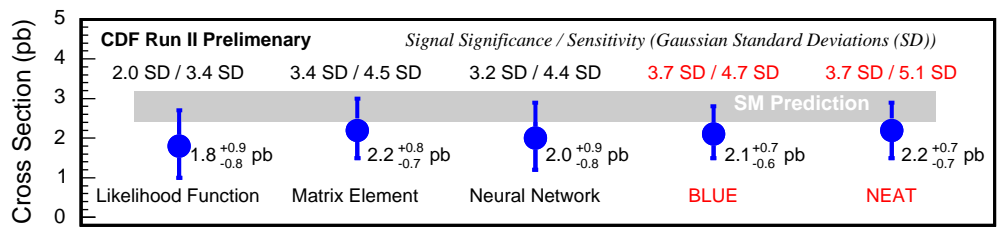}
  \caption{Cross section measurements and sensitivity results for each analysis and the BLUE and NEAT combinations using $2.2~fb^{-1}$ of CDF data.} \label{fig:results}
\end{figure*}

The results of the the combination and each individual analysis are
summarized in Figure~\ref{fig:results}.  Fitting for the combined
single-top cross section NEAT measures $\sigma_{s+t} = 2.2 \pm 0.7$
pb which corresponds to a measurement of
$|V_{tb}|=0.88\pm0.14(\textrm{exp})\pm0.07(\textrm{theory})$.  In addition the NEAT
combination improves the expected sensitivity ($p$-value~\cite{pdg})by about 9~\%.  The BLUE
cross check measures a consistent cross section with NEAT and gains about 7~\%
in expected sensitivity.  In addition, BLUE was used to study the
consistency of the three analyses and finds that the combined measurement
had a $\chi^{2}$ which was better than 87~\% of
pseudo-experiments. Also, about 15~\% of pseudo-experiments thrown with the standard
model expectation of single top events measured a cross section below
2.1~pb.  So, BLUE confirms that the three CDF measurements are highly
compatible with one another and that the combined result represents a
deviation from the standard model expectation of about 1$\sigma$.

In summary, the three CDF analyses have been combined with two
different techniques resulting in an improved sensitivity and
measurement of the single top production cross section.

\end{document}